\documentclass[aps,prb,superscriptaddress,twocolumn]{revtex4-2}
\usepackage[T1]{fontenc}
\usepackage{graphicx}
\usepackage{xcolor}
\usepackage{csquotes}
\usepackage{nameref}
\usepackage{hyperref}
\usepackage{amsmath, amssymb, amsthm, amsfonts, mathtools}
\usepackage{natbib}
\usepackage{geometry}
\newcommand{\Bayreuth}{Theoretische Physik III, Universit{\"a}t Bayreuth, 95440 Bayreuth, Germany}
\newcommand{\Dortmund}{Condensed Matter Theory, Department of Physics, TU Dortmund, 44221 Dortmund, Germany}
\newcommand{\Vienna}{Vienna Center for Quantum Science and Technology, University of Vienna, 1090 Vienna, Austria}

\begin{document}
\title{Photon Number Coherence of a Quantum Dot-Cavity System Excited Using the SUPER Scheme}

\author{Paul C. A. Hagen}
\email{paul.hagen@uni-bayreuth.de}
\affiliation{\Bayreuth}

\author{Thomas Bracht}
\affiliation{\Dortmund}

\author{Mathieu Bozzio}
\affiliation{\Vienna}

\author{Moritz Cygorek}
\affiliation{\Dortmund}

\author{Doris E. Reiter}
\affiliation{\Dortmund}

\author{Philip Walther}
\affiliation{\Vienna}

\author{Vollrath M. Axt}
\affiliation{\Bayreuth}

\begin{abstract}
    To fulfill the security requirements of quantum cryptography, photon number coherence (PNC) of single photon sources has recently become an important figure of merit. Quantum dots (QDs) embedded in photonic microcavities offer a mature source of single photons, of which many properties can be tuned by the use of different excitation protocols or parameters. We show that the \enquote{Swing-UP of quantum EmitteR population} (SUPER) scheme can significantly decrease the PNC of the emitted photon, compared to resonant excitation. The reason for this is a laser-induced Stark shift, which effectively decouples the QD from the cavity during the SUPER excitation. Our calculations account for environmental effects such as phonons and radiative losses.
\end{abstract}

\maketitle

Single photon sources play a key role in the implementation of quantum communications~\cite{Hao-Ze2025, Liu2023}. Besides the typical figures of merit --single photon purity~\cite{Kiraz2004, Cosacchi2019, Bozzio2022, Michler2000, Giesz2016, Wei2014}, indistinguishability~\cite{Wei2014, Sbresny2022, Vajner2024} and brightness~\cite{Cosacchi2019, Bozzio2022, Michler2000, Vajner2024}-- the photon number coherence (PNC) has emerged as crucial quantity for quantum cryptographic protocols~\cite{Bozzio2022, Karli2024}. PNC refers to the coherences between photonic Fock states. In the case of single-photon sources, the PNC is dominated by the coherence between the vacuum and the single-photon Fock state~\cite{Hagen2025}. Typically, protocols such as BB84/decoy quantum key distribution (QKD) \cite{Bozzio2022, Karli2024, LoPreskill2007, Dusek2000} require small PNC for their security proofs to hold, while some such as twin field QKD~\cite{Bozzio2022} or weak coin flipping~\cite{Bozzio2020}, benefit from large PNC. Most investigations of PNC have so far been focused on sources with Poissonian photon statistics~\cite{LoPreskill2007, Dusek2000} and only recently has this been expanded to more general sources~\cite{Hagen2025}.\\

Like other figures of merit, the PNC can strongly depend on the protocol used to excite the single-photon source. For example in the case of resonant excitation, the PNC can be tuned by varying the pulse area~\cite{Hagen2025}. However, resonant excitation is not always optimal for practical applications, e.g. because the emitted photons have the same energy as scattered photons from the excitation beam, requiring challenging cross-polarization filtering \cite{Somaschi2016, Senellart2017}. On the other hand, spectral separation between excitation and emission can be realized using different driving schemes, such as phonon-assisted state preparation \cite{Glaessl2013, Cosacchi2019} or off-resonant two-color schemes \cite{Yan2025, Boos2024}. One of these is the \enquote{Swing-up of quantum EmitteR population} scheme~\cite{Bracht2021, Bracht2023a}, which realizes population inversion using two red-detuned laser pulses. Intuitively, one laser can be understood to dress the system, while the other laser drives resonant transitions between the laser dressed states~\cite{Bracht2023a}.\\

When SUPER is applied to a QD in a cavity, the Stark shift induced by the laser dressing, temporarily detunes the QD from the cavity during the laser excitation~\cite{Heinisch2024, Bracht2023b}. Hence, photons are effectively only emitted into the cavity at the end of the laser pulse. Because this suppresses the emission into the cavity during the excitation, which is known to be the reason for sizable PNC under resonant excitation, SUPER can be expected to allow for smaller PNC than resonant excitation. Indeed, we find that SUPER can produce much smaller PNC than resonant excitation, while allowing for similar tunability. This provides an interesting alternative to standard PNC scrambling methods used in quantum cryptography, such as discrete phase modulation~\cite{Cao2015} and laser gain switching~\cite{Kobayashi2014}, which present known security loopholes~\cite{Curras-Loredo2023}.\\

In this article, we therefore explore the impact of the SUPER excitation scheme on PNC. To this end, we employ a model consisting of a QD, a quantized microcavity, phonons and radiative as well as cavity losses.

\section{Photon Number Coherence}
We begin by formulating the density matrix for a photonic state consisting of zero or one photons:
\begin{equation}
        \rho = \left(\begin{array}{cc}
        \rho_{00}& |\rho_{01}|\,\text{e}^{\text{i}\phi} \\
         |\rho_{01}|\,\text{e}^{-\text{i}\phi} & \rho_{11}
    \end{array}\right),
\end{equation}
where $\rho_{00}$ and $\rho_{11}$ are the vacuum and single photon populations, respectively, and $\rho_{01} = |\rho_{01}|\,\text{e}^{\text{i}\phi}$ is the PNC between the zero- and one-photon states, fulfilling
\begin{equation}
    |\rho_{01}|^2 \leq \rho_{00}\rho_{11}.
\end{equation}
In case of equality, the state is pure, otherwise it is mixed.\\

Ref.~\cite{Loredo2019} has shown how PNC can be measured: A photon emitted from the single-photon source is interfered in a Hong-Ou-Mandel setup~\cite{Hong1987} with a delayed photon from the same source, triggered by an earlier pulse. Then the intensity at the output ports $c$ and $d$ of the HOM beamsplitter is predicted to be~\cite{Loredo2019}
\begin{equation} \label{eq:HOM_Correlations_mixed}
    \mathcal{N}_{c,d} = \rho_{11}\left(1\pm \frac{|\rho_{01}|^2}{\rho_{11}}\,\text{cos}(\phi^a - \phi^b)\right).
\end{equation}
Using Eq.~\eqref{eq:HOM_Correlations_mixed}, it is possible to measure the magnitude of the PNC $|\rho_{01}|$, if the single-photon population $\rho_{11}$ is known~\cite{Loredo2019, Karli2024}.\\

Concretely, as a figure of merit, we use the time-integrated absolute value
\begin{equation}
    \tilde{\rho}_{01} = \int_{-\infty}^\infty |\rho_{01}(t)|\,\text{d}t
\end{equation}
and indicate time-integration by the tilde $\tilde{.}$.

\section{QD-Cavity System}
\begin{figure}
    \centering
    \includegraphics[scale=0.85]{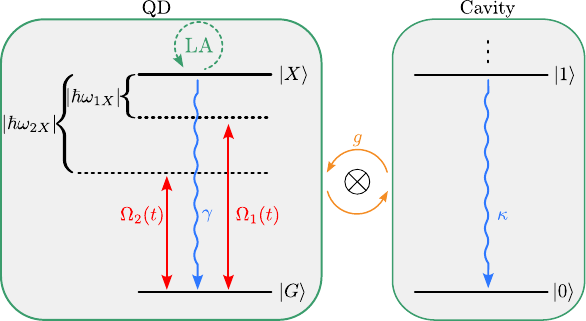}
    \caption{Level scheme of the QD-cavity system that is excited using the SUPER scheme and couples to longitudinal acoustic phonons. The lasers $\Omega_{1/2}(t)$ are both red-detuned from the exciton-transition by $\hbar\omega_{1/2\,X}$. The QD couples to the cavity through the coupling strength $g$ and can decay with a rate $\gamma$, while the cavity decay rate is $\kappa$. The exciton also couples to the environment of longitudinal acoustic (LA) phonons.}
    \label{fig:levels}
\end{figure}
The QD-cavity system is schematically displayed in Fig.~\ref{fig:levels}. It consists of the ground state $|G\rangle$ and an excited state $|X\rangle$ of the QD, which couples to a cavity, that is represented by Fock states $|n\rangle$ of a single cavity mode with $n\in\mathbb{N}_0$. In our numerical calculations, we include $n\leq 2$. This is justified since under our excitation conditions, already the $n=2$ Fock state is less occupied than the $n=1$ Fock state by at least two orders of magnitude for all times.\\

The SUPER scheme makes use of two lasers, which are red-detuned from the resonance frequency of the two-level system. Their instantaneous Rabi frequencies are displayed in Fig.~\ref{fig:levels} as $\Omega_{1/2}$ and their detunings to the exciton are $\hbar\omega_{1/2\,X}$. The sign convention is such that $\hbar\omega_{1/2\,X} < 0\text{ meV}$ refers to red-detuned lasers. We will refer to the higher-energy laser as the \enquote{first} laser, and the lower-energy one will be the \enquote{second} laser.\\

We choose SUPER parameters that produce simultaneously high brightness $\tilde{\rho}_{11}$ and low PNC $\tilde{\rho}_{01}$. In particular, we used $\hbar\omega_{1X} = -2.06$~meV, $\Theta_1 = 15.95\,\pi$, $\hbar\omega_{2X} = -5.08$~meV $\Theta_2 = 15.78\,\pi$. Both laser envelopes are Gaussian
\begin{equation}
    \Omega_i(t) = \frac{\Theta_i}{\sqrt{2\pi}\sigma}\text{e}^{-\frac{t^2}{2\sigma^2}}
\end{equation}
and their durations $\sigma$ are chosen to be identical. $\Theta_i$ are the respective pulse areas with $i\in\{1,2\}$.\\

For self-assembled InGaAs/GaAs QDs it is important to include the influence of longitudinal acoustic phonons, which are the main source of dephasing~\cite{Wang1998}. Their influence is included through the use of a pure-dephasing Hamiltonian~\cite{Krummheuer2002}. The QD can spontaneously decay back into the ground state, which we include using a decay rate $\gamma$. Similarly, the emission out of the cavity is described using a rate $\kappa$. We use numerically exact methods to solve the von Neumann equation~\cite{Cygorek2022}. More details about the system and the numerical methods used for solving it can be found in Appendix A.\\


\begin{figure*}
    \centering
    \includegraphics[width=\linewidth]{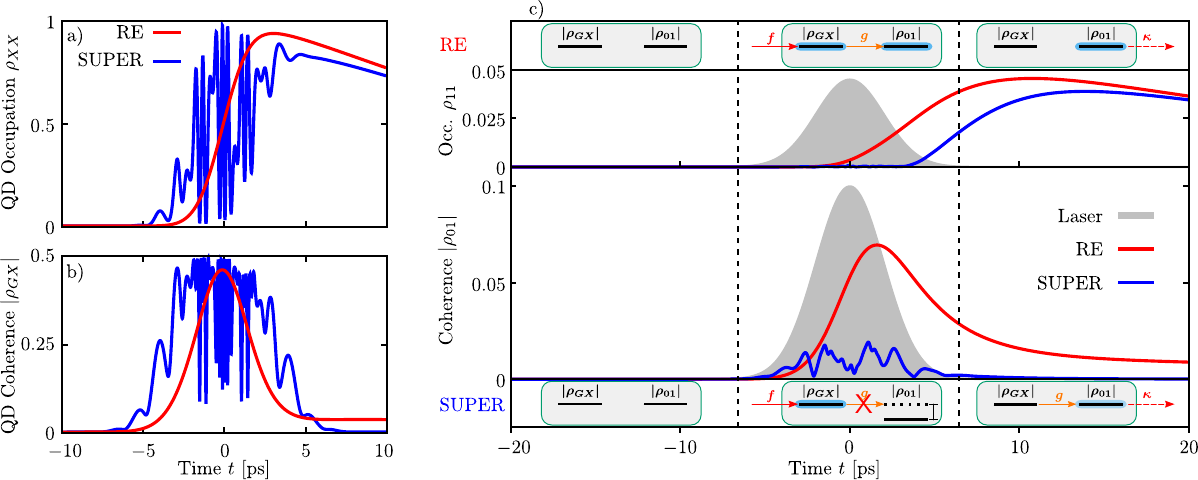}
    \caption{a) Shows the QD occupation for resonant excitation (red) and SUPER (blue) in dependence on time. b) shows the respective absolute electronic coherences. The cavity occupations and coherences between 0 and 1 photons in the cavity are shown in c), together with schematics illustrating the physical processes occurring for RE and SUPER, respectively. Both SUPER pulses and the RE pulse have identical durations with $\sigma = 2$~ps. The RE pulse area was chosen by minimizing $\tilde{\rho}_{01} - \tilde{\rho}_{11}$ to make sure low coherence and high occupation is achieved. All calculations were done at 4.2~K.}
    \label{fig:centralResults}
\end{figure*}

\section{Dynamics during the excitation}
The QD population and electronic coherence of the calculations done for resonant excitation (RE) and SUPER are displayed in Fig.~\ref{fig:centralResults}~a) and ~b). All pulses have the same duration $\sigma = 2$~ps and the delay between the SUPER pulses is zero, to allow for easier comparison. This relatively short pulse duration was chosen because it ensures that RE performs relatively well, since it produces smaller PNC for shorter pulses.~\cite{Hagen2025}\\

For both excitation schemes, the occupation increases during the pulse until it reaches a maximum at the end of the pulse and then decreases due to radiative decay and the coupling to the cavity. The most notable difference between RE and SUPER is that SUPER additionally to the above trends also displays quick oscillations due to its off-resonant nature.\\

In Fig.~\ref{fig:centralResults}~b) we see that the electronic coherence also shows similar trend behavior for RE and SUPER. Both initially increase until half-way through the pulse and then decrease again. This decrease, can easily be understood by recalling that electronic coherences in general have to fulfill
\begin{equation} \label{eq:elecCohLimit}
    |\rho_{GX}|^2 \leq \rho_{GG}\,\rho_{XX},
\end{equation}
where the right-hand side is maximal for equal ground-state and excited-state populations $\rho_{GG}$ and $\rho_{XX}$, i.e. for $\rho_{GG} = \rho_{XX} = 0.5$. As a consequence, coherent excitation schemes, like RE and SUPER, have their highest electronic coherences at the time of half-excitation, which occurs half-way through the pulse. Afterwards, as a high population of the exciton is reached, the coherence gets more restrained by Eq.~\eqref{eq:elecCohLimit} and therefore decreases as the QD gets closer to the exciton state.\\

In Ref.~\cite{Hagen2025} it was found that the electronic coherence, and with it PNC, is increased for RE, if phonons are present. This coherence-boosting mechanism does not exist for SUPER anymore and they instead decrease both, the occupation and the coherence, which is shown in Appendix B. In Fig.~\ref{fig:centralResults}~a) and b), this can be seen in the SUPER results simultaneously showing lower $\rho_{XX}$ and $|\rho_{GX}|$ than RE for $t > 5$~ps.\\

The respective occupations and coherences of the cavity Fock states are displayed in Fig.~\ref{fig:centralResults}~c) for RE and SUPER. Note that under the given excitation conditions only the states $|0\rangle$ and $|1\rangle$ acquire notably non-zero occupations. Considering the cavity occupation, one observes that the emission from the QD into the cavity occurs later for SUPER than it does for RE. For RE, emission into the cavity occurs during the excitation. In contrast, the emission for SUPER takes place just after the excitation has ended. Consequently, the photonic coherence is much larger during the pulse for RE than it is for SUPER. In the latter case, the build-up of photonic coherence is not shifted to times after the pulse has passed. In contrast to RE, there is never a large amount of coherence within the cavity if the SUPER is used.\\

We attribute both of these phenomena to the Stark shift, which SUPER exerts on the QD \cite{Bracht2023b, Heinisch2024}. It is schematically displayed by the drawings at the top and bottom of Fig.~\ref{fig:centralResults}~c). For both excitation schemes, the QD and the cavity are initially in resonance. When one shines resonant light at this system, there is no Stark-shift and therefore both, the occupation and the coherence can simply be emitted into the cavity (see top, central schematic) and is afterwards outcoupled from the cavity. For SUPER, however, the Stark-Shift induced by this excitation scheme, effectively detunes the QD from the cavity, prohibiting emission. As a consequence, neither the cavity occupation, nor its coherence rise as the pulses are active. Consequently, the QD is still very highly excited after the laser pulses passed and little occupation has been emitted into the cavity. Because at this point, the QD is so close to the excited state for either excitation protocol, Eq.~\eqref{eq:elecCohLimit} states that the electronic coherence must be quite small as well, as we have elaborated earlier. In the following, we investigate what consequences this has for the photonic coherence.\\

For all excitation protocols, for which the QD-cavity system behaves as a single-photon source, $\rho_{01}$ originates the same way~\cite{Hagen2025}. In this case, the photonic coherence $\rho_{01}$ and the electronic coherence $\rho_{GX}$ are connected via~\cite{Hagen2025}
\begin{equation}
    \rho_{01}(t) = -\text{i} g \int_{-\infty}^t\rho_{GX}(t')\,\text{e}^{\left(-\text{i}\omega_{CX}-\frac{\kappa}{2}\right)(t-t')}\,\text{d}t',
\end{equation}
where we assumed to be in a frame rotating with the excitonic frequency $\omega_X$. As we can see, the photonic coherence is influenced by all past electronic coherences, which are convoluted with a memory kernel. This kernel, the exponential function within the integral, constitutes a Lorentzian filter, which only transmits if the coherence $\rho_{GX}$ oscillates with a frequency that matches $\omega_{CX}$. Transfer into the cavity is expected for $\hbar\omega_{CX}\approx 0$~meV, i.e. the cavity should be resonant to the QD transition. Also, the occupations $\rho_{XX}$ or $\rho_{11}$ do not enter the behavior of the coherence. Because of this, one finds that a perfectly excited QD, which according to Eq.~\eqref{eq:elecCohLimit} has $\rho_{GX}=0$, would not produce any PNC whatsoever.\\

For SUPER, the emission into the cavity can not take place while the lasers are active, because the QD is Stark shifted out of resonance with the cavity by the first laser. The QD is almost fully excited by the time the lasers have passed. The electronic coherence is small by the time the QD and cavity can couple effectively. This little residual coherence is then converted into PNC and can be seen in Fig.~\ref{fig:centralResults}~c) as the very slight increase in $|\rho_{01}|$ after the pulse. \\

Resulting from those different excitation dynamics, we find the following PNC for SUPER and RE: $\tilde{\rho}_{01}^{\text{SUPER}} = 0.10~\text{ps}$ and $\tilde{\rho}_{01}^{\text{RE}} = 1.14~\text{ps}$.
This shows that, in this case, simply using SUPER instead of RE decreases PNC by about 90\% when the pulse durations of all lasers are identical.

\section{Controlling PNC with SUPER} \label{sec:controllingPNCWithSUPER}
\begin{figure}
    \centering
    \includegraphics[scale=1]{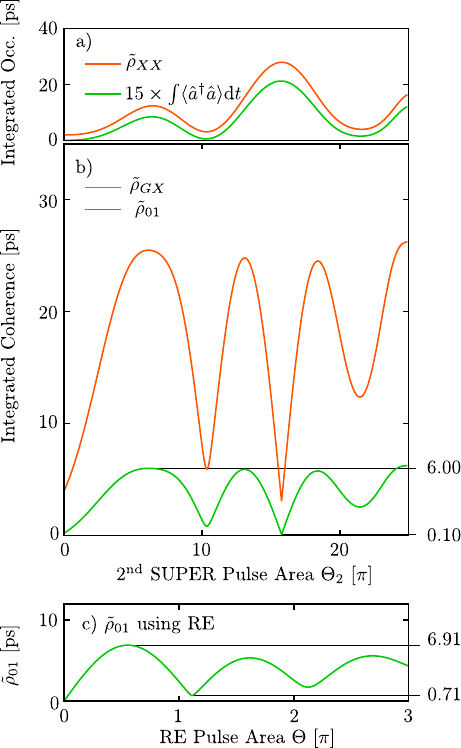}
    \caption{Integrated values of a) the occupations and b) the coherences for the QD (orange) and the cavity photons (green) in dependence on the second SUPER pulse area, while the first pulse area is kept constant. c) Shows the PNC, which is obtained for resonant excitation with a pulse of equal duration $\sigma = 2$~ps to the SUPER pulses in a) and b).}
    \label{fig:rabis}
\end{figure}

Ref.~\cite{Bracht2023a} found that SUPER can be understood as the second laser producing resonant excitation between the first laser's dressed states. Because of this, we vary only the second pulse area $\Theta_2$, leaving all other parameters the same as before. This results in only partial excitation in the system, which has a significant effect on the coherences. The resulting integrated occupations $\tilde{\rho}_{XX}$ and $\int\langle\hat{a}^\dagger\hat{a}\rangle\,\text{d}t$ are displayed in Fig.~\ref{fig:rabis}~a), where Rabi-like behavior is clearly visible.\\

The integrated coherences $\tilde{\rho}_{GX}$ and $\tilde{\rho}_{01}$ are shown in Fig.~\ref{fig:rabis}~b). The coherences also show the well-known Rabi-rotations~\cite{Hagen2025}, with one exception: The fist coherence maximum lies at roughly the same pulse area as the first occupation maximum, which is explained by the small occupation equal to roughly half the other maxima heights. For higher pulse areas, the occupations and coherences follow the more familiar pattern, in which the coherence is minimal whenever the occupation has an extremum. Interestingly, the electronic coherence is noticeably non-zero for $\Theta_2 = 0\,\pi$. This is due to far off-resonant excitations that the first laser can produce. Indeed, the integrated QD occupation displayed in Fig.~\ref{fig:rabis}~a) is also slightly non-zero in this case. For comparison, the PNC created using a RE pulse of identical duration to our SUPER pulses is displayed in Fig.~\ref{fig:rabis}~c).

These results show that the SUPER scheme possesses the ability to tune the PNC by changing the second pulse area, while leaving the first laser unchanged. In fact, SUPER can achieve significantly lower PNC values (0.10 ps) than RE (0.71 ps), which is beneficial for many quantum-cryptographic protocols where the PNC should be erased~\cite{Bozzio2022}. On the other hand, the maximal PNC values achievable using SUPER (6.00 ps)in Fig.~\ref{fig:rabis} are slightly smaller than those of RE (6.91 ps), while the width of the tunable interval stays roughly the same.\\

\section{Conclusion}
We showed for a QD-cavity model including the effect of phonons that the SUPER scheme can produce single photons with significantly smaller PNC than a resonant pulse of equal duration could, while still allowing a large degree of tunability by variation of the second pulse area. This large tunability implies that QDs excited with SUPER are very versatile: They can be used both in quantum-cryptographic protocols that require PNC such as twin-field QKD~\cite{Bozzio2022} and quantum weak coin flipping~\cite{Bozzio2020} or in protocols that require no PNC such as BB84/decoy QKD and some mistrustful quantum primitives~\cite{Bozzio2022}.\\

The reason for this is the effective decoupling from the cavity during the excitation, which occurs because of the Stark shift induced by the first laser. This shift allows for the excitation to stay in the QD until the end of the laser driving, such that the electronic coherence decreases as the QD approaches the excited state. Additionally, if short pulses are used, SUPER excitation triggers an incoherent phonon-mediated mechanism, decreasing the PNC even further, at the cost of some brightness. At the end of the driving, SUPER leaves only a small electronic coherence in the QD that is converted to only a small PNC. During the pulse, the conversion is suppressed because of the effective decoupling of the cavity. In contrast, RE continuously builds up PNC, particularly while the QD is excited half-way and thus while the electronic coherence is maximal.\\


\section*{Appendix A: The QD-cavity system}
SUPER can be used to excite any two-level system. For this publication, we specifically consider a QD consisting of a ground state $|G\rangle$ and an excited state $|X\rangle$. Transitions between these states are described by the operators $\hat{\sigma} = |G\rangle\langle X|$ and $\hat{\sigma}^\dagger = |X\rangle\langle G|$. SUPER requires driving the system by two temporally overlapping pulses, both of which are red detuned from the exciton transition.\\

The Hamiltonian of the two-level system and the higher-frequency laser, which we refer to as the \enquote{first} laser from here on, is given by
\begin{equation}
    \hat{H}_1 = -\frac{\hbar\Omega_1(t)}{2}\left(\text{e}^{-\text{i}\omega_{1X}t}\hat{\sigma}^\dagger + \text{e}^{\text{i}\omega_{1X}t}\hat{\sigma}\right),
\end{equation}
where the rotating frame was chosen resonant to the $|G\rangle\leftrightarrow|X\rangle$ transition, from which $\omega_{1X}$ describes the detuning of the first laser.  Similarly, the second laser is described by
\begin{equation}
    \hat{H}_2 = -\frac{\hbar\Omega_2(t)}{2}\left(\text{e}^{-\text{i}\omega_{2X} t}\hat{\sigma}^\dagger + \text{e}^{\text{i}\omega_{2X} t}\hat{\sigma}\right).
\end{equation}
$\omega_{2X}$ is the frequency difference between the second laser and the exciton. $\Omega_2(t)$ is the pulse envelope of the second laser.
Both laser envelopes are Gaussian
\begin{equation}
    \Omega_i(t) = \frac{\Theta_i}{\sqrt{2\pi}\sigma}\text{e}^{-\frac{t^2}{2\sigma^2}}
\end{equation}
and their durations $\sigma$ are chosen to be identical. $\Theta_i$ are the respective pulse areas.\\

We place the QD inside of a photonic cavity, which we model as a single photonic mode. The single photonic mode is described by the Jaynes-Cummings Hamiltonian~\cite{Nazir2016}
\begin{equation}
    \hat{H}_{JC} = \hbar\omega_{CX}\,\hat{a}^\dagger\hat{a} + \hbar g(\hat{a}\hat{\sigma}^\dagger + \hat{a}^\dagger\hat{\sigma}),
\end{equation}
where $\hat{a}$ ($\hat{a}^\dagger$) are the annihilation (creation) operators and $\hbar\omega_{CX}$ is the detuning between the cavity and the exciton. We choose for the coupling constant $g=0.05$~meV~\cite{Hopfmann2015, Kistner2010, Yoshie2004} and include up to two photons in our calculation. In this paper, we only discuss the case with $\hbar\omega_{CX}=0$~meV because this strongly facilitates the conversion of dot excitation into cavity photons. The coupling of the QD to its longitudinal-acoustic phonon environment is taken into account through the Hamiltonian
\begin{equation}
    \hat{H}_{\text{ph}} = \hbar\sum_j\omega_j\hat{b}_j^\dagger\hat{b}_j + \hbar \hat{\sigma}^\dagger\hat{\sigma}\sum_j\left(\gamma_j\hat{b}^\dagger_j + \gamma_j^*\hat{b}_j\right),
\end{equation}
where the sum runs over all phonon modes, which have the annihilation (creation) operators $\hat{b}_j$ ($\hat{b}^\dagger_j$) and energy $\hbar\omega_j$. The coupling of the $j$th mode to the exciton is determined by $\gamma_j$. We choose the same phonon spectral density
\begin{equation}
    J(\omega) = \sum_j|\gamma_j|^2\delta(\omega-\omega_j)
\end{equation}
as in Refs.~\cite{Hagen2025, Barth2016}, which fully characterizes all phononic influences~\cite{Barth2016, Leggett1987, MakriMakarov1994}. The radius of the electronic density-distribution in the QD is chosen to be $a_e = 3$~nm.\\

The full Hamiltonian is thus given by
\begin{equation} \label{eq:Hamiltonian}
    \hat{H} = \hat{H}_1 + \hat{H}_2 + \hat{H}_{JC} + \hat{H}_{\text{ph}}.
\end{equation}
Additionally, we take losses into account via the Lindblad operators \cite{Lindblad1976, Carmichael1993}
\begin{equation}
    \mathcal{L}_{\hat{O},\delta}[\hat{\rho}] = \delta\left(\hat{O}\hat{\rho}\hat{O}^\dagger - \frac{1}{2}\left[\hat{\rho},\hat{O}^\dagger\hat{O}\right]_+\right),
\end{equation}
where $\hat{O}$ is an operator, $\delta$ is a rate and $[\,\cdot\,,\,\cdot\,]_+$ denotes the anti-commutator. We consider Lindbladians to describe the radiative decay $\mathcal{L}_{\hat{\sigma},\gamma}$ and cavity losses $\mathcal{L}_{\hat{a},\kappa}$. With these contributions, the model is described by a Liouville-von Neumann equation
\begin{equation} \label{eq:vonNeumannEquation}
    \frac{\text{d}}{\text{d}t} \hat{\rho}(t) = -\frac{\text{i}}{\hbar}[\hat{H}(t),\hat{\rho}(t)] + \mathcal{L}_{\hat{\sigma},\gamma}\left[\hat{\rho}(t)\right] + \mathcal{L}_{\hat{a},\kappa}\left[\hat{\rho}(t)\right].
\end{equation}
The level scheme of the full model is drawn in Fig.~\ref{fig:levels}.

We solve Eq.~\eqref{eq:vonNeumannEquation} using the process tensor matrix product operators~\cite{JoergensenPollock2019} implemented in the Automatic Compression of Environments (ACE) code~\cite{Cygorek2022, Cygorek2024a, Cygorek2024b}, which allows for the numerically exact inclusion of the phonon environment, while also being able to resolve very short-time behavior on the scale of $\sim$10~fs. This is necessary due to the quick oscillations that occur solving SUPER, because of the large detunings involved. ACE thus provides us with a reduced density matrix $\rho_{Sn}^{S'n'}$, with $S,S'\in\{G,X\}$ and $n,n'\in\mathbb{N}_0$, describing the QD-cavity system. The effect of phonons on the QD-cavity system is fully included, but the phononic degrees of freedom are removed using the trace over the phononic subspace $\text{Tr}_\text{ph}$.\\

To obtain the reduced density matrix describing only the cavity, the QD degrees of freedom are traced out
\begin{align} \label{eq:TraceOutQD}
        \rho_{nm}(t) \coloneqq \sum_{S\in\{G,X\}} \text{Tr}_\text{ph}\left[\hat{\rho}(t)\,|Sn\rangle\langle Sm|\right] && n,m\in\mathbb{N}_0,
\end{align}
where $|Sn\rangle = |S\rangle\otimes|n\rangle$, and to obtain the reduced density matrix describing the QD, the cavity is traced out
\begin{equation}
    \begin{split}
        \rho_{SS'}(t) \coloneqq \sum_{n\in\mathbb{N}_0} \text{Tr}_\text{ph}\left[\hat{\rho}(t)\,|Sn\rangle\langle S'n|\right]\hspace{1pt} S,S'\in\{G,X\}.
    \end{split}
\end{equation}
This gives access to the QD and photon occupations $\rho_{XX}$ and $\rho_{nn}$, respectively, as well as the electronic coherence $\rho_{GX}$ and photonic coherences $\rho_{nm}$, $n\neq m$.\\

\section{Appendix B: Phononic influence}
\begin{figure*}
    \centering
    \includegraphics[scale=1]{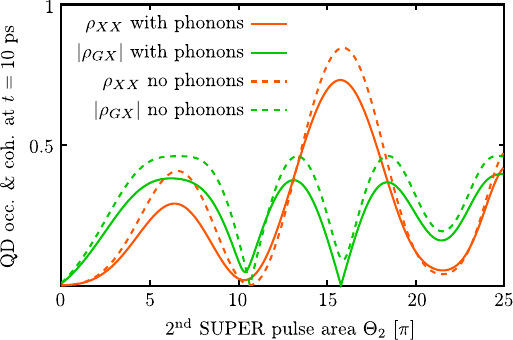}
    \caption{The electronic occupation $\rho_{XX}$ and the absolute value of the coherence $|\rho_{Gx}|$ at time 10~ps after the pulse maximum. The solid (dashed) lines show the results with (without) phonons. Fig.~\ref{fig:centralResults} uses $\Theta_2 = 15.78~\pi$.}
    \label{fig:checkPhonLDSTransit}
\end{figure*}

We have observed in Fig.~\ref{fig:centralResults}~a) and b) that the occupation and the coherence after the excitation finished, are both smaller for SUPER than they are for RE. If the excitation was perfectly coherent, this could not occur, because Eq.~\eqref{eq:elecCohLimit} would increase the coherence as the occupation runs towards 0.5. This implies that there are additional incoherent effects that become relevant in our system. The most probable of these are the phonons, which can have different effects on the system, depending on the excitation protocol. Fig.~\ref{fig:checkPhonLDSTransit} shows the electronic occupation $\rho_{XX}$ and coherence $\rho_{GX}$ just after the pulse ended at time $t=10$~ps with and without phonons. The decay rates and the cavity are included. The phonon-free occupation reaches essentially perfect occupation, when no decay or cavity is present (not shown), so the apparently small occupation in the phonon-free case in Fig.~\ref{fig:checkPhonLDSTransit} is due to those emissions.\\

Fig.~\ref{fig:checkPhonLDSTransit} confirms that the decrease in occupation and coherence is due to phononic effects. Notice that those effects occur practically independently from $\Theta_2$. In particular, this shows that the coherence-boosting effect phonons can have for resonant excitation~\cite{Hagen2025}, is not universal and does not occur for SUPER.

\bibliography{references}

@article{Hao-Ze2025,
    author = {Chen, Hao-Ze and Li, Ming-Han and Wang, Yu Zhou and Zhao, Zhen-Geng and Ye, Cheng and Li, Fei Long and Chen, Zhu and Han, Sheng-Long and Tang, Bao and Miao, Ya Jun and Qi, Wei},
    title = {Implementation of carrier-grade quantum communication networks over 10000 km},
    journal = {npj Quantum Information},
    year = {2025},
    volume = {11},
    number = {1},
    pages = {137},
    doi = {10.1038/s41534-025-01089-8},
    url = {https://doi.org/10.1038/s41534-025-01089-8}
}

@article{Liu2023,
  title = {Experimental Twin-Field Quantum Key Distribution over 1000 km Fiber Distance},
  author = {Liu, Yang and Zhang, Wei-Jun and Jiang, Cong and Chen, Jiu-Peng and Zhang, Chi and Pan, Wen-Xin and Ma, Di and Dong, Hao and Xiong, Jia-Min and Zhang, Cheng-Jun and Li, Hao and Wang, Rui-Chun and Wu, Jun and Chen, Teng-Yun and You, Lixing and Wang, Xiang-Bin and Zhang, Qiang and Pan, Jian-Wei},
  journal = {Phys. Rev. Lett.},
  volume = {130},
  issue = {21},
  pages = {210801},
  numpages = {6},
  year = {2023},
  month = {May},
  publisher = {American Physical Society},
  doi = {10.1103/PhysRevLett.130.210801},
  url = {https://link.aps.org/doi/10.1103/PhysRevLett.130.210801}
}

@article{Kiraz2004,
  title = {Quantum-dot single-photon sources: Prospects for applications in linear optics quantum-information processing},
  author = {Kiraz, A. and Atat\"ure, M. and Imamo\ifmmode \breve{g}\else \u{g}\fi{}lu, A.},
  journal = {Phys. Rev. A},
  volume = {69},
  issue = {3},
  pages = {032305},
  numpages = {10},
  year = {2004},
  month = {Mar},
  publisher = {American Physical Society},
  doi = {10.1103/PhysRevA.69.032305},
  url = {https://link.aps.org/doi/10.1103/PhysRevA.69.032305}
}

@article{Cosacchi2019,
  title = {Emission-Frequency Separated High Quality Single-Photon Sources Enabled by Phonons},
  author = {Cosacchi, M. and Ungar, F. and Cygorek, M. and Vagov, A. and Axt, V. M.},
  journal = {Phys. Rev. Lett.},
  volume = {123},
  issue = {1},
  pages = {017403},
  numpages = {5},
  year = {2019},
  month = {Jul},
  publisher = {American Physical Society},
  doi = {10.1103/PhysRevLett.123.017403},
  url = {https://link.aps.org/doi/10.1103/PhysRevLett.123.017403}
}

@article{Bozzio2020,
  title = {Quantum weak coin flipping with a single photon},
  author = {Bozzio, Mathieu and Chabaud, Ulysse and Kerenidis, Iordanis and Diamanti, Eleni},
  journal = {Phys. Rev. A},
  volume = {102},
  issue = {2},
  pages = {022414},
  numpages = {15},
  year = {2020},
  month = {Aug},
  publisher = {American Physical Society},
  doi = {10.1103/PhysRevA.102.022414},
  url = {https://link.aps.org/doi/10.1103/PhysRevA.102.022414}
}

@article{Bozzio2022,
    author = {Bozzio, Mathieu and Vyvlecka, Michal and Cosacchi, Michael and Nawrath, Cornelius and Seidelmann, Tim and Loredo, Juan C. and Portalupi, Simone L. and Axt, Vollrath M. and Michler, Peter and Walther, Philip},
    title = {Enhancing quantum cryptography with quantum dot single-photon sources},
    journal = {npj Quantum Information},
    year = {2022},
    volume = {8},
    number = {1},
    pages = {104},
    doi = {10.1038/s41534-022-00626-z},
    url = {https://doi.org/10.1038/s41534-022-00626-z}
}

@article{Michler2000,
    author = {P. Michler  and A. Kiraz  and C. Becher  and W. V. Schoenfeld  and P. M. Petroff  and Lidong Zhang  and E. Hu  and A. Imamoglu },
    title = {A Quantum Dot Single-Photon Turnstile Device},
    journal = {Science},
    volume = {290},
    number = {5500},
    pages = {2282-2285},
    year = {2000},
    doi = {10.1126/science.290.5500.2282},
    URL = {https://www.science.org/doi/abs/10.1126/science.290.5500.2282}
}

@article{Giesz2016,
    author = {Giesz, V. and Somaschi, N. and Hornecker, G. and Grange, T. and Reznychenko, B. and De Santis, L. and Demory, J. and Gomez, C. and Sagnes, I. and Lemaître, A. and Krebs, O. and Lanzillotti-Kimura, N. D. and Lanco, L. and Auffeves, A. and Senellart, P.},
    title = {Coherent manipulation of a solid-state artificial atom with few photons},
    journal = {Nature Communications},
    year = {2016},
    volume = {7},
    number = {1},
    pages = {11986},
    doi = {10.1038/ncomms11986},
    url = {https://doi.org/10.1038/ncomms11986}
}

@article{Wei2014,
    author = {Wei, Yu-Jia and He, Yu-Ming and Chen, Ming-Cheng and Hu, Yi-Nan and He, Yu and Wu, Dian and Schneider, Christian and Kamp, Martin and Höfling, Sven and Lu, Chao-Yang and Pan, Jian-Wei},
    title = {Deterministic and Robust Generation of Single Photons from a Single Quantum Dot with 99.5{\%} Indistinguishability Using Adiabatic Rapid Passage},
    journal = {Nano Letters},
    year = {2014},
    volume = {14},
    number = {11},
    pages = {6515--6519},
    doi = {10.1021/nl503081n},
    url = {https://doi.org/10.1021/nl503081n}
}

@article{Sbresny2022,
  title = {Stimulated Generation of Indistinguishable Single Photons from a Quantum Ladder System},
  author = {Sbresny, Friedrich and Hanschke, Lukas and Sch\"oll, Eva and Rauhaus, William and Scaparra, Bianca and Boos, Katarina and Zubizarreta Casalengua, Eduardo and Riedl, Hubert and del Valle, Elena and Finley, Jonathan J. and J\"ons, Klaus D. and M\"uller, Kai},
  journal = {Phys. Rev. Lett.},
  volume = {128},
  issue = {9},
  pages = {093603},
  numpages = {7},
  year = {2022},
  month = {Mar},
  publisher = {American Physical Society},
  doi = {10.1103/PhysRevLett.128.093603},
  url = {https://link.aps.org/doi/10.1103/PhysRevLett.128.093603}
}

@article{Vajner2024,
    author = {Vajner, Daniel A. and Holewa, Paweł and Zięba-Ostój, Emilia and Wasiluk, Maja and von Helversen, Martin and Sakanas, Aurimas and Huck, Alexander and Yvind, Kresten and Gregersen, Niels and Musiał, Anna and Syperek, Marcin and Semenova, Elizaveta and Heindel, Tobias},
    title = {On-Demand Generation of Indistinguishable Photons in the Telecom C-Band Using Quantum Dot Devices},
    journal = {ACS Photonics},
    year = {2024},
    volume = {11},
    number = {2},
    pages = {339--347},
    doi = {10.1021/acsphotonics.3c00973},
    url = {https://doi.org/10.1021/acsphotonics.3c00973}
}

@INPROCEEDINGS{Wang1998,
    author={Hailin Wang and Xudong Fan and Takagahara, T. and Cunningham, J.E.}, booktitle={Technical Digest. Summaries of Papers Presented at the International Quantum Electronics Conference. Conference Edition. 1998 Technical Digest Series, Vol.7 (IEEE Cat. No.98CH36236)},
    title={Pure dephasing induced by exciton-phonon interactions in GaAs quantum dots},
    year={1998},
    volume={},
    number={},
    pages={162-163},
    doi={10.1109/IQEC.1998.680336}
}

@article{Krummheuer2002,
  title = {Theory of pure dephasing and the resulting absorption line shape in semiconductor quantum dots},
  author = {Krummheuer, B. and Axt, V. M. and Kuhn, T.},
  journal = {Phys. Rev. B},
  volume = {65},
  issue = {19},
  pages = {195313},
  numpages = {12},
  year = {2002},
  month = {May},
  publisher = {American Physical Society},
  doi = {10.1103/PhysRevB.65.195313},
  url = {https://link.aps.org/doi/10.1103/PhysRevB.65.195313}
}

@article{Glaessl2013,
  title = {Proposed Robust and High-Fidelity Preparation of Excitons and Biexcitons in Semiconductor Quantum Dots Making Active Use of Phonons},
  author = {Gl\"assl, M. and Barth, A. M. and Axt, V. M.},
  journal = {Phys. Rev. Lett.},
  volume = {110},
  issue = {14},
  pages = {147401},
  numpages = {5},
  year = {2013},
  month = {Apr},
  publisher = {American Physical Society},
  doi = {10.1103/PhysRevLett.110.147401},
  url = {https://link.aps.org/doi/10.1103/PhysRevLett.110.147401}
}

@article{Nazir2016,
doi = {10.1088/0953-8984/28/10/103002},
url = {https://dx.doi.org/10.1088/0953-8984/28/10/103002},
year = {2016},
month = {feb},
publisher = {IOP Publishing},
volume = {28},
number = {10},
pages = {103002},
author = {Nazir, Ahsan and McCutcheon, Dara P S},
title = {Modelling exciton–phonon interactions in optically driven quantum dots},
journal = {Journal of Physics: Condensed Matter}
}

@article{Kistner2010,
    author = {Kistner, C. and Morgener, K. and Reitzenstein, S. and Schneider, C. and Höfling, S. and Worschech, L. and Forchel, A. and Yao, P. and Hughes, S.},
    title = {Strong coupling in a quantum dot micropillar system under electrical current injection},
    journal = {Applied Physics Letters},
    volume = {96},
    number = {22},
    pages = {221102},
    year = {2010},
    month = {06},
    issn = {0003-6951},
    doi = {10.1063/1.3442912},
    url = {https://doi.org/10.1063/1.3442912}
}

@article{Hopfmann2015,
  title = {Compensation of phonon-induced renormalization of vacuum Rabi splitting in large quantum dots: Towards temperature-stable strong coupling in the solid state with quantum dot-micropillars},
  author = {Hopfmann, C. and Musia\l{}, A. and Strau\ss{}, M. and Barth, A. M. and Gl\"assl, M. and Vagov, A. and Strau\ss{}, M. and Schneider, C. and H\"ofling, S. and Kamp, M. and Axt, V. M. and Reitzenstein, S.},
  journal = {Phys. Rev. B},
  volume = {92},
  issue = {24},
  pages = {245403},
  numpages = {10},
  year = {2015},
  month = {Dec},
  publisher = {American Physical Society},
  doi = {10.1103/PhysRevB.92.245403},
  url = {https://link.aps.org/doi/10.1103/PhysRevB.92.245403}
}

@article{Yoshie2004,
    author = {Yoshie, T. and Scherer, A. and Hendrickson, J. and Khitrova, G. and Gibbs, H. M. and Rupper, G. and Ell, C. and Shchekin, O. B. and Deppe, D. G.},
    title = {Vacuum Rabi splitting with a single quantum dot in a photonic crystal nanocavity},
    journal = {Nature},
    year = {2004},
    volume = {432},
    number = {7014},
    pages = {200--203},
    doi = {10.1038/nature03119},
    url = {https://doi.org/10.1038/nature03119}
}

@article{Somaschi2016,
    author = {Somaschi, N. and Giesz, V. and De Santis, L. and Loredo, J. C. and Almeida, M. P. and Hornecker, G. and Portalupi, S. L. and Grange, T. and Antón, C. and Demory, J. and Gómez, C. and Sagnes, I. and Lanzillotti-Kimura, N. D. and Lemaítre, A. and Auffeves, A. and White, A. G. and Lanco, L. and Senellart, P.}, 
    title = {Near-optimal single-photon sources in the solid state},
    journal = {Nature Photonics},
    volume = {10},
    number = {5},
    pages = {340--345},
    year = {2016},
    doi = {10.1038/nphoton.2016.23},
    url = {https://doi.org/10.1038/nphoton.2016.23}
}

@article{Senellart2017,
    author = {Senellart, Pascale and Solomon, Glenn and White, Andrew},
    title = {High-performance semiconductor quantum-dot single-photon sources},
    journal = {Nature Nanotechnology},
    year = {2017},
    volume = {12},
    number = {11},
    pages = {1026--1039},
    doi = {10.1038/nnano.2017.218},
    url = {https://doi.org/10.1038/nnano.2017.218}
}

@misc{Yan2025,
      title={Robust entangled photon generation enabled by single-shot Floquet driving}, 
      author={Jun-Yong Yan and Paul C. A. Hagen and Hans-Georg Babin and Wei E. I. Sha and Andreas D. Wieck and Arne Ludwig and Chao-Yuan Jin and Vollrath M. Axt and Da-Wei Wang and Moritz Cygorek and Feng Liu},
      year={2025},
      eprint = {2504.02753},
      archivePrefix={arXiv},
      primaryClass={quant-ph},
      url={https://arxiv.org/abs/2504.02753}, 
}

@article{Bracht2021,
  title = {Swing-Up of Quantum Emitter Population Using Detuned Pulses},
  author = {Bracht, Thomas K. and Cosacchi, Michael and Seidelmann, Tim and Cygorek, Moritz and Vagov, Alexei and Axt, V. Martin and Heindel, Tobias and Reiter, Doris E.},
  journal = {PRX Quantum},
  volume = {2},
  issue = {4},
  pages = {040354},
  numpages = {11},
  year = {2021},
  month = {Dec},
  publisher = {American Physical Society},
  doi = {10.1103/PRXQuantum.2.040354},
  url = {https://link.aps.org/doi/10.1103/PRXQuantum.2.040354}
}

@article{Bracht2023a,
  title = {Dressed-state analysis of two-color excitation schemes},
  author = {Bracht, Thomas K. and Seidelmann, Tim and Karli, Yusuf and Kappe, Florian and Remesh, Vikas and Weihs, Gregor and Axt, Vollrath Martin and Reiter, Doris E.},
  journal = {Phys. Rev. B},
  volume = {107},
  issue = {3},
  pages = {035425},
  numpages = {11},
  year = {2023},
  month = {Jan},
  publisher = {American Physical Society},
  doi = {10.1103/PhysRevB.107.035425},
  url = {https://link.aps.org/doi/10.1103/PhysRevB.107.035425}
}

@article{Bracht2023b,
author = {Thomas K. Bracht and Moritz Cygorek and Tim Seidelmann and Vollrath Martin Axt and Doris E. Reiter},
journal = {Optica Quantum},
number = {2},
pages = {103--107},
publisher = {Optica Publishing Group},
title = {Temperature-independent almost perfect photon entanglement from quantum dots via the SUPER scheme},
volume = {1},
month = {Dec},
year = {2023},
url = {https://opg.optica.org/opticaq/abstract.cfm?URI=opticaq-1-2-103},
doi = {10.1364/OPTICAQ.498559},
}

@article{Heinisch2024,
  title = {Swing-up dynamics in quantum emitter cavity systems: Near ideal single photons and entangled photon pairs},
  author = {Heinisch, Nils and K\"ocher, Nikolas and Bauch, David and Schumacher, Stefan},
  journal = {Phys. Rev. Res.},
  volume = {6},
  issue = {1},
  pages = {L012017},
  numpages = {7},
  year = {2024},
  month = {Jan},
  publisher = {American Physical Society},
  doi = {10.1103/PhysRevResearch.6.L012017},
  url = {https://link.aps.org/doi/10.1103/PhysRevResearch.6.L012017}
}

@article{Boos2024,
author = {Boos, Katarina and Sbresny, Friedrich and Kim, Sang Kyu and Kremser, Malte and Riedl, Hubert and Bopp, Frederik W. and Rauhaus, William and Scaparra, Bianca and Jöns, Klaus D. and Finley, Jonathan J. and Müller, Kai and Hanschke, Lukas},
title = {Coherent Swing-Up Excitation for Semiconductor Quantum Dots},
journal = {Advanced Quantum Technologies},
volume = {7},
number = {4},
pages = {2300359},
doi = {https://doi.org/10.1002/qute.202300359},
url = {https://advanced.onlinelibrary.wiley.com/doi/abs/10.1002/qute.202300359},
eprint = {https://advanced.onlinelibrary.wiley.com/doi/pdf/10.1002/qute.202300359},
year = {2024}
}

@article{Hagen2025,
author = {Hagen, Paul C. A. and Bozzio, Mathieu and Cygorek, Moritz and Reiter, Doris E. and Axt, Vollrath M.},
title = {Photon Number Coherence in Quantum Dot-Cavity Systems can be Enhanced by Phonons},
journal = {Advanced Quantum Technologies},
volume = {8},
number = {7},
pages = {2400455},
doi = {https://doi.org/10.1002/qute.202400455},
url = {https://advanced.onlinelibrary.wiley.com/doi/abs/10.1002/qute.202400455},
year = {2025}
}

@article{Karli2024,
    author = {Karli, Yusuf and Vajner, Daniel A. and Kappe, Florian and Hagen, Paul C. A. and Hansen, Lena M. and Schwarz, René and Bracht, Thomas K. and Schimpf, Christian and Covre da Silva, Saimon F. and Walther, Philip and Rastelli, Armando and Axt, Vollrath Martin and Loredo, Juan C. and Remesh, Vikas and Heindel, Tobias and Reiter, Doris E. and Weihs, Gregor},
    title = {Controlling the photon number coherence of solid-state quantum light sources for quantum cryptography},
    journal = {npj Quantum Information},
    year = {2024},
    volume = {10},
    number = {1},
    pages = {17},
    doi = {10.1038/s41534-024-00811-2},
    url = {https://doi.org/10.1038/s41534-024-00811-2}
}

@article{Dusek2000,
  title = {Unambiguous state discrimination in quantum cryptography with weak coherent states},
  author = {Du\ifmmode \check{s}\else \v{s}\fi{}ek, Miloslav and Jahma, Mika and L\"utkenhaus, Norbert},
  journal = {Phys. Rev. A},
  volume = {62},
  issue = {2},
  pages = {022306},
  numpages = {9},
  year = {2000},
  month = {Jul},
  publisher = {American Physical Society},
  doi = {10.1103/PhysRevA.62.022306},
  url = {https://link.aps.org/doi/10.1103/PhysRevA.62.022306}
}

@article{LoPreskill2007,
  author       = {Hoi{-}Kwong Lo and
                  John Preskill},
  title        = {Security of quantum key distribution using weak coherent states with
                  nonrandom phases},
  journal      = {Quantum Inf. Comput.},
  volume       = {7},
  number       = {5},
  pages        = {431--458},
  year         = {2007},
  doi          = {10.26421/QIC7.5-6-2},
  url       = {https://dblp.org/rec/journals/qic/LoP07.bib}
}

@article{Loredo2019,
    author = {Loredo, J. C. and Antón, C. and Reznychenko, B. and Hilaire, P. and Harouri, A. and Millet, C. and Ollivier, H. and Somaschi, N. and De Santis, L. and Lemaître, A. and Sagnes, I. and Lanco, L. and Auffèves, A. and Krebs, O. and Senellart, P.},
    title = {Generation of non-classical light in a photon-number superposition},
    journal = {Nature Photonics},
    year = {2019},
    volume = {13},
    number = {11},
    pages = {803--808},
    doi = {10.1038/s41566-019-0506-3},
    url = {https://doi.org/10.1038/s41566-019-0506-3}
}

@article{Hong1987,
  title = {Measurement of subpicosecond time intervals between two photons by interference},
  author = {Hong, C. K. and Ou, Z. Y. and Mandel, L.},
  journal = {Phys. Rev. Lett.},
  volume = {59},
  issue = {18},
  pages = {2044--2046},
  numpages = {0},
  year = {1987},
  month = {Nov},
  publisher = {American Physical Society},
  doi = {10.1103/PhysRevLett.59.2044},
  url = {https://link.aps.org/doi/10.1103/PhysRevLett.59.2044}
}

@article{Lindblad1976,
	Author = {Lindblad, G. },
	Da = {1976/06/01},
	Doi = {10.1007/BF01608499},
	Id = {Lindblad1976},
	Isbn = {1432-0916},
	Journal = {Comm. Math. Phys.},
	Pages = {119--130},
	Title = {On the generators of quantum dynamical semigroups},
	Ty = {JOUR},
	Url = {https://doi.org/10.1007/BF01608499},
	Volume = {48},
	Year = {1976}
}

@book{Carmichael1993,
    author = {H. Carmichael},
    title = {Lecture Notes in Physics: An Open Systems Approach to Quantum Optics},
    publisher = {Springer},
    address = {Heidelberg},
    year = {1993}
}

@article{Barth2016,
  title = {Path-integral description of combined Hamiltonian and non-Hamiltonian dynamics in quantum dissipative systems},
  author = {Barth, A. M. and Vagov, A. and Axt, V. M.},
  journal = {Phys. Rev. B},
  volume = {94},
  issue = {12},
  pages = {125439},
  numpages = {9},
  year = {2016},
  month = {Sep},
  publisher = {American Physical Society},
  doi = {10.1103/PhysRevB.94.125439},
  url = {https://link.aps.org/doi/10.1103/PhysRevB.94.125439}
}

@article{MakriMakarov1994,
title = {Path integrals for dissipative systems by tensor multiplication. Condensed phase quantum dynamics for arbitrarily long time},
journal = {Chemical Physics Letters},
volume = {221},
number = {5},
pages = {482-491},
year = {1994},
doi = {https://doi.org/10.1016/0009-2614(94)00275-4},
author = {Dmitrii E. Makarov and Nancy Makri}
}

@article{Leggett1987,
  title = {Dynamics of the dissipative two-state system},
  author = {Leggett, A. J. and Chakravarty, S. and Dorsey, A. T. and Fisher, Matthew P. A. and Garg, Anupam and Zwerger, W.},
  journal = {Rev. Mod. Phys.},
  volume = {59},
  issue = {1},
  pages = {1--85},
  numpages = {0},
  year = {1987},
  month = {Jan},
  publisher = {American Physical Society},
  doi = {10.1103/RevModPhys.59.1},
  url = {https://link.aps.org/doi/10.1103/RevModPhys.59.1}
}

@article{JoergensenPollock2019,
title = {Exploiting the Causal Tensor Network Structure of Quantum Processes to Efficiently Simulate Non-Markovian Path Integrals},
  author = {J\o{}rgensen, Mathias R. and Pollock, Felix A.},
  journal = {Phys. Rev. Lett.},
  volume = {123},
  issue = {24},
  pages = {240602},
  numpages = {7},
  year = {2019},
  month = {Dec},
  publisher = {American Physical Society},
  doi = {10.1103/PhysRevLett.123.240602},
  url_disabled = {https://link.aps.org/doi/10.1103/PhysRevLett.123.240602}
}

@article{Cygorek2022,
    author = {Cygorek, Moritz and Cosacchi, Michael and Vagov, Alexei and Axt, Vollrath Martin and Lovett, Brendon W. and Keeling, Jonathan and Gauger, Erik M.},
    title = {Simulation of open quantum systems by automated compression of arbitrary environments},
    journal = {Nature Physics},
    year = {2022},
    volume = {18},
    number = {6},
    pages = {662--668},
    doi = {10.1038/s41567-022-01544-9},
    url_disabled = {https://doi.org/10.1038/s41567-022-01544-9}
}

@article{Cygorek2024a,
  title = {Sublinear Scaling in Non-Markovian Open Quantum Systems Simulations},
  author = {Cygorek, Moritz and Keeling, Jonathan and Lovett, Brendon W. and Gauger, Erik M.},
  journal = {Phys. Rev. X},
  volume = {14},
  issue = {1},
  pages = {011010},
  numpages = {25},
  year = {2024},
  month = {Feb},
  publisher = {American Physical Society},
  doi = {10.1103/PhysRevX.14.011010},
  url_disabled= {https://link.aps.org/doi/10.1103/PhysRevX.14.011010}
}

@article{Cygorek2024b,
    author = {Cygorek, Moritz and Gauger, Erik M.},
    title = {ACE: A general-purpose non-Markovian open quantum systems simulation toolkit based on process tensors},
    journal = {The Journal of Chemical Physics},
    volume = {161},
    number = {7},
    pages = {074111},
    year = {2024},
    month = {08},
    issn = {0021-9606},
    doi = {10.1063/5.0221182},
    url_disabled = {https://doi.org/10.1063/5.0221182},
}

@article{Cao2015,
doi = {10.1088/1367-2630/17/5/053014},
url = {https://doi.org/10.1088/1367-2630/17/5/053014},
year = {2015},
month = {may},
publisher = {IOP Publishing},
volume = {17},
number = {5},
pages = {053014},
author = {Cao, Zhu and Zhang, Zhen and Lo, Hoi-Kwong and Ma, Xiongfeng},
title = {Discrete-phase-randomized coherent state source and its application in quantum key distribution},
journal = {New Journal of Physics}
}

@article{Kobayashi2014,
  title = {Evaluation of the phase randomness of a light source in quantum-key-distribution systems with an attenuated laser},
  author = {Kobayashi, Toshiya and Tomita, Akihisa and Okamoto, Atsushi},
  journal = {Phys. Rev. A},
  volume = {90},
  issue = {3},
  pages = {032320},
  numpages = {9},
  year = {2014},
  month = {Sep},
  publisher = {American Physical Society},
  doi = {10.1103/PhysRevA.90.032320},
  url = {https://link.aps.org/doi/10.1103/PhysRevA.90.032320}
}

@article{Curras-Loredo2023,
doi = {10.1088/2058-9565/ad141c},
url = {https://doi.org/10.1088/2058-9565/ad141c},
year = {2023},
month = {dec},
publisher = {IOP Publishing},
volume = {9},
number = {1},
pages = {015025},
author = {Currás-Lorenzo, Guillermo and Nahar, Shlok and Lütkenhaus, Norbert and Tamaki, Kiyoshi and Curty, Marcos},
title = {Security of quantum key distribution with imperfect phase randomisation},
journal = {Quantum Science and Technology}
}
\end{document}